\documentclass[11pt]{article}
\usepackage{graphicx}
\usepackage{amssymb}

\newtheorem{proposition}{Proposition}

\def\proof{\noindent  {\underline {Proof}}. }
\def\square{ {\hfill \vrule height6pt width6pt depth1pt} \bigskip \medskip }

\begin{document}

\centerline{\Large Equations in dual variables for Whittaker functions.}

\vskip 1cm
\centerline{O. Babelon
\footnote{Member of CNRS.}}
\vskip .5cm
\centerline{Laboratoire de Physique Th\'eorique et Hautes Energies.
\footnote{L.P.T.H.E. Universit\'es Paris VI--Paris VII (UMR 7589),
 Bo\^{\i}te 126, Tour 16, $1^{er}$ \'etage,
 4 place Jussieu, F-75252 PARIS CEDEX 05}
}
\date{May  2003}

\vskip 5cm

\begin{abstract}
It is known that the Whittaker functions $w(q,\lambda)$ 
associated to the group $SL(N)$ are  eigenfunctions
of the Hamiltonians of the open Toda chain, hence satisfy a set of 
differential equations in the Toda variables $q_i$.
Using the expression of the $q_i$ for the closed Toda chain in terms of
Sklyanin variables $\lambda_i$, and the known relations 
between the open and the closed Toda chains, we show that 
Whittaker functions  
also satisfy a set of  new difference equations in $\lambda_i$. 
\end{abstract}

\vfill
LPTHE-03-22
\eject

\section{Introduction.}

The method of separation of variables is acquiring a central
place in the domain of integrable systems. In the classical case,
the separated variables are simply the poles of the eigenvectors
of the Lax matrix. Together with the spectral curve, 
they are the necessary data to reconstruct the eigenvectors
from their analyticity properties, and therefore
the Lax matrix itself.
The solution of the integrable system is given the fact
that the image of the divisor of poles by the Abel map evolves linearly on the Jacobian under an integrable flow.

In the quantum case, these separated variables where first defined 
and used by Sklyanin. In particular, he showed that the original 
$N$-body Shroedinger equation separates into $N$ one dimensional
equations, called Baxter equations. 
The separated variables also appear in the $N$-soliton form factor formulae
in sine-Gordon theory. It was shown that Smirnov formulae have a simple
interpretation in terms of the separated variables of the $N$ solitons
\cite{BBS96}.

More recently, it was realized that the separated variables provided a 
convenient approach to the problem of quantizing a classical 
integrable system. It is interesting that this idea emerged
 independently both among mathematicians \cite{ER01} and physicists   \cite{BaTa02}.     

In \cite{Ba03}, we took this point of view to solve
 the quantum inverse problem for the closed Toda chain.
 In this note, we remark that this result immediately suggests
 that the Whittaker functions associated to $Sl(N)$, which are 
 the eigenfunctions of the open Toda chain, see eqs.(\ref{kostant}) below, should also satisfy
 a set of equations in the dual (momentum) variables, 
 eqs.(\ref{xkeq3}) below.
 We prove these relations by using the Mellin-Barnes integral
 representation for the Whittaker functions found in \cite{KhLe00}.

\section{The Closed Toda Chain.}

The closed Toda chain is defined by the Hamiltonian

\begin{equation}
H =  \sum_{i=1}^{N+1}{1\over 2} p_i^2 + e^{q_{i+1}-q_i}
\label{hamtoda}
\end{equation}
where we assume that $q_{N+2} \equiv q_1$, and canonical Poisson
brackets
$$
\{q_i,q_j\}=0,\quad \{p_i,q_j\} = \delta_{ij},\quad \{p_i,p_j\}=0
$$

As it is well known, one can associate to this system
a $2\times 2$ Lax matrix as follows. Consider the matrices
$$
T_j(\lambda) = \pmatrix{\lambda + p_j  & -e^{q_j} \cr e^{-q_j} &0}
$$
and construct
\begin{equation}
T(\lambda) = T_{1}(\lambda)\cdots T_2(\lambda)T_{N+1}(\lambda)
\label{nparticles}
\end{equation}
We can write
\begin{equation}
T(\lambda) = \pmatrix{A(\lambda) & B(\lambda) \cr
                                  C(\lambda) & D(\lambda) }; \quad A(\lambda) D(\lambda) - B(\lambda) C(\lambda) =1
\label{TToda}
\end{equation}
$A(\lambda)$ is a polynomial of degree $N+1$, $D(\lambda)$ is  of degree $N-1$, and
$B(\lambda)$, $C(\lambda)$ are of degree $n$.
The spectral curve is defined as usual
\begin{equation}
\det (T(\lambda) - \mu) = 0 \quad  \equiv \quad \mu + \mu^{-1} - t(\lambda) = 0
\label{spectralcurve}
\end{equation}
where 
$$
t(\lambda)=A(\lambda) + D(\lambda) = \lambda^{N+1} + \sum_{j=0}^{N} \lambda^j H_j, \quad
H_{N} = P,\quad H_{N-1} = {1\over 2} P^2 - H
$$
where $P = \sum_i p_i$, and $H$ is given by eq.(\ref{hamtoda}). 
We will consider the system reduced by the translational symmetry. We 
fix the total momentum  $P=0$. The symplectic form $\omega = \sum_{i=1}^{N+1} d p_i\wedge dq_i $ becomes
$\omega_{reduced} = \sum_{i=1}^N dp_i \wedge d(q_i-q_{N+1})$ so that the canonical coordinates  of the reduced system can be taken as 
$(p_i, q_i-q_{N+1})$, $i=1\cdots N$. We choose the
gauge condition $q_{N+1}=0$. It must be emphasized  that this reduced system, which does not contain the degree of freedom $p_{N+1},q_{N+1}$ anymore,
 is not the open Toda chain.
The $N$ quantities $H_j$ are conserved and Poisson commute.

The separated variables are the poles of the eigenvectors of the Lax matrix. 
For a $2\times 2$ matrix of the form eq.(\ref{TToda})
the eigenvector is simple
$$
(T(\lambda) - \mu) \Psi = 0, \quad \Psi = \pmatrix{1\cr \psi_2 }, \quad \psi_2 = -{A(\lambda) - \mu \over B(\lambda)}
$$
The poles of $\psi_2$ at finite distance are above the zeroes  $\lambda_i$ of $B(\lambda)=0$ which is a polynomial
of degree $n$. The two points above $\lambda_i$ are $\mu_i^+ = A(\lambda_i)$, $\mu_i^- = D(\lambda_i)$ so that 
$\psi_2$ has a pole only at the second point. The points of the dynamical divisor are therefore $(\lambda_i, D(\lambda_i)), \quad B(\lambda_i) = 0$.

The Inverse Problem consists in reconstructing everything 
 in terms of the $(\lambda_i,\mu_i)$.  
Its solution for the quantum theory was given in \cite{Ba03}.

Quantum commutation relations are defined directly on the separated variables.
$$
[\lambda_k,\lambda_{k'}] =0,\quad \mu_k \lambda_{k'} = (\lambda_{k'} + i\hbar\delta_{kk'} ) \mu_k,\quad
[\mu_k,\mu_{k'}] =0
$$

To reconstruct  the Hamiltonians
themselves is simple.
The $n$ points $\lambda_i,\mu_i^-$ belonging to the spectral curve, we have the equations (we drop the superscript $-$ in $\mu_i^-$):
$$
\mu_i + \mu_i^{-1} - \lambda_i^{N+1} - \sum_{j=1}^{N} \lambda_i^j H_j = 0
$$
This is a linear system of $N$ equations for the $N$ quantities $H_j$.
Its solution makes sense even quantum mechanically and, quite 
generally, produces commuting Hamiltonians \cite{ER01,BaTa02}.

To reconstruct  the original operators $p_i,q_i$,  we 
need the operators $X^{(k)}, Y^{(k)}$ below.
We call $[k]$ a subset of cardinality $k$ of  $(1,2,\cdots, N)$: $
[k]= (i_1,i_2,\cdots , i_k)$. 
We write $\sum_{[k]}$ for the sum over all such subsets. Let
\begin{equation}
S_{[k]}(\lambda) = \prod_{i\in [k]}\prod_{j\not \in [k]} {1\over (\lambda_{i} -\lambda_j)}, \quad {\rm and} \quad
 \mu_{[k]} = \mu_{i_1}\mu_{i_2}\cdots \mu_{i_k}
\label{defSk}
\end{equation}
and define
\begin{eqnarray}
X^{(k)}&\equiv&   \sum_{[k]}   S_{[k]}(\lambda) \mu_{[k]}
\label{defXr} \\
Y^{(k)}&\equiv&  
\sum_{[k]} S_{[k]} (\lambda)
\left( \sum_{i\not\in [k]} \lambda_i \right)  \mu_{[k]} 
\label{defYr}
\end{eqnarray}
The quantum Toda operators are given by:
\begin{equation}
e^{q_i} =(-1)^{N-1}{X^{(N-i+1)}\over X^{(N-i)}},\quad p_i = {Y^{(N-i+1)}\over X^{(N-i+1)}}-{Y^{(N-i)}\over X^{(N-i)}}
\label{reconquant}
\end{equation}
The canonical commutation relations
\begin{equation}
\left[e^{q_i} , e^{q_j} \right] =0,\quad 
\left[ e^{q_i} , p_j\right] =
- i\hbar \delta_{ij}e^{q_i},\quad
 \left[ p_i, p_j \right] = 0
 \label{canquant} 
\end{equation}
are a simple consequences of the following quadratic algebra:
\begin{eqnarray*}
[ X^{(k)},X^{(l)}] &=& 0 \\
 \relax [ X^{(k)},Y^{(l)}] &=&  i\hbar (k-l)  \theta (k-l) X^{(k)} X^{(l)}   \\
\relax [ Y^{(k)},Y^{(l)}] &=&i\hbar (k-l) \Big[ \theta(k-l) Y^{(k)} X^{(l)}+\theta(l-k) Y^{(l)} X^{(k)} \Big] 
\end{eqnarray*}
Note that this implies that there is no ordering ambiguity in the expressions eq.(\ref{reconquant}). 

The operators $X^{(k)}$ are self adjoint with respect to the non trivial
scalar product
$$
(f,g) = \int_{-\infty}^\infty d \lambda \; m(\lambda)\; f^*(\lambda) g(\lambda)
$$
where
\begin{equation}
m(\lambda) = \prod_{i < j} {1Ê\over \left\vert \Gamma ({\lambda_i - \lambda_j \over i\hbar })\right\vert^2}
\label{measure}
\end{equation}

\section{Whittaker vectors and functions.}

There is an important connection between the closed and open Toda chains that we now recall.

The matrix element $B(\lambda)$ in eq.(\ref{TToda})
can be written as
$$
B(\lambda) =-e^{q_{N+1}} \sum_{k=0}^N  \lambda^{N-k} h_k(p,q)
$$
where $h_k(p,q)$ are the Hamiltonians of the {\em open} Toda chain
obtained by removing particle $N+1$ from the closed chain \cite{Gut81}. 

The eigenfunctions of the
Hamiltonians $h_k(p,q)$ satisfy
\begin{equation}
h_k(p,q) w(\lambda,q) = \sigma_k(\lambda) w(\lambda,q)
\label{kostant}
\end{equation}
where $\sigma_k(\lambda)$ are the symmetric functions of the 
$\lambda_i$. 
It is a beautiful result by Kostant and Semenov-Tian-Shansky that $w(\lambda,q)$ is the Whittaker function of $sl(N)$.
Thus the function $w(\lambda,q)$ is the kernel of the Fourier
 transform going from  the $q$
to the  $\lambda$ (momentum) variables. 
They satisfy the completness relation
\begin{equation}
\int d\lambda \; m(\lambda) w^*(q,\lambda) w(q',\lambda) =
  \delta(q-q')  \label{completness}
\end{equation}
where $m(\lambda)$ is given by eq.(\ref{measure}), 
and the orthogonality relations \cite{KhLe99}
\begin{equation}
\int dq \; w^*(q,\lambda) w(q,\lambda') =  m^{-1}(\lambda) 
\delta(\lambda-\lambda') \label{orthogonality}
\end{equation}

To recall how such functions arise, consider the $sl(N)$ Lie algebra
$$
[H_j,E_{\pm \alpha_i}] = \pm a_{ij}E_{\pm \alpha_i},\quad 
[E_{ \alpha_i},E_{- \alpha_j}] = \delta_{ij}H_i,\quad i,j=1,\cdots ,N-1
$$
where $\alpha_i$ are the simple root vectors and $a_{ij}$ is the 
Cartan matrix. The Weyl vector is
$$
\rho = {1\over 2} \sum_{\alpha > 0}\alpha
$$
The quadratic Casimir operator is
$$
C_2 = \sum_{\alpha} E_\alpha E_{-\alpha} + \sum_{i,j}a^{-1}_{ij} H_i H_j
$$
where the $\sum_\alpha$ runs over all positive and negative roots.

Whittaker vectors are such that
\begin{equation}
E_{\alpha_i} \vert w_\lambda \rangle =  \mu_{\alpha_i}^R  \vert w_\lambda \rangle,\quad \alpha_i ~~{\rm simple}
\label{whittakerright}
\end{equation}
Of course, if $\alpha$ is a non-simple root, then $E_{\alpha} \vert w_\lambda \rangle = 0$.
Similarly, dual Whittaker vectors satisfy
\begin{eqnarray}
 \langle w_\lambda  \vert E_{-\alpha_i} &=&  \mu_{\alpha_i}^L   \langle w_\lambda  \vert,\quad \alpha_i ~~{\rm simple}
 \label{whittakerleft}
\end{eqnarray}
We may furthermore assume that $\vert w_\lambda \rangle $ belongs to some 
irreducible representation with weight $\lambda$. In that case, we have
$$
C_2 \vert w_\lambda \rangle =  c_2(\lambda)  \vert w_\lambda \rangle
$$
Define the Whittaker function
$$
w(\lambda,q) = e^{\rho(q)} \langle w_\lambda  \vert e^{q}\vert w_\lambda \rangle
$$
where $q= \sum_{ij} q_i a_{ij}^{-1} H_j$ belongs to the Cartan subalgebra.
Then, we have
\begin{eqnarray*}
c_2(\lambda)w(\lambda,q)&=&e^{\rho(q)}
\langle w_\lambda  \vert e^{q }C_2\vert w_\lambda \rangle \\
&=&e^{\rho(q)}\langle w_\lambda \vert e^{q }\left(
2\sum_{\alpha>0} E_{-\alpha}E_\alpha  + 2 H_\rho + \sum_{i,j}a^{-1}_{ij} H_i H_j
\right)\vert w_\lambda \rangle
\end{eqnarray*}
Expanding $H_\alpha = \sum_i (\alpha)_i a_{ij}^{-1}H_j$,
where $a_{ij}$ is the Cartan matrix, we get
$$
c_2(\lambda)w(\lambda,q) = \left( a_{ij}{\partial^2\over \partial q_i \partial q_j} 
 + 2 \sum_{\alpha\; {\rm simple}} \mu_\alpha^{L} \mu_\alpha^R e^{-\alpha(q) } 
 \right) w(\lambda,q)
$$
The differential operator in the right hand side is just the Hamiltonian
of the open Toda chain.
Hence $ w(\lambda,q)$ satisfy to the open Toda chain Schroedinger equation! Clearly, the same analysis can be 
done for higher order Casimirs giving rise to the higher order Hamiltonians
of the open chain.

There exists several integral representations for the Whittaker functions.
\begin{itemize}
\item {\bf The "Gauss" representation} \cite{GKMMMO95}.

$$
w_G(\lambda,q) = \int dz_{ij} \prod_{i=1}^{N-1} \Delta_i^{-(\lambda, \alpha_i) -1}
(zS^{-1}) e^{\sum_i \mu_i^R e^{\alpha_i(q)}z_{i,i+1} + \mu_{N-i}^L 
{\Delta_{i,i+1}(zS^{-1}) \over \Delta_i(zS^{-1}) }}
$$
The integral is over upper-triangular matrices $N\times N$ matrices, $z$, 
with $1$ on the diagonal. $S$ is the antidiagonal matrix
$S_{ij} = \delta_{N+1-i,j}$.
 $\Delta_i(M)$ is the determinant of the $i\times i$ submatrix of 
$M$ consisting in the first $i$ rows and columns.  $\Delta_{i-1,i}(M)$ is defined as the determinant of the 
$(i-1)\times (i-1)$ submatrix of $M$ with columns $i-1$ and $i$ interchanged.

\item {\bf The "Iwasawa" representation} \cite{GKMMMO95}.

$$
w_I(\lambda,q) = e^{\lambda(q)} \int dz_{ij} \prod_{i=1}^{N-1} \Delta_i^{-{1\over 2}(\lambda, \alpha_i) -{1\over 2}}( zz^t) e^{i \sum_i \mu_i z_{i,i+1} e^{\alpha_i(q)}}
$$
The definitions of $z$ and $\Delta_i(M)$ are the same as above.

\item {\bf The "Mellin-Barnes" representation} \cite{KhLe00}.

The Weyl invariant Whittaker function has a representation in terms of multiple Mellin-Barnes integrals.
Let $\gamma_{jk}$ be a lower triangular $N\times N$ matrix.
$$
\pmatrix{\gamma_{11} & 0 & \cdots  & & \cr
\gamma_{21}& \gamma_{22} & 0 & &\cdots \cr 
\vdots & & & \cr
\gamma_{N-1,1} & \gamma_{N-1,2} & \cdots & \gamma_{N-1,N-1} & 0 \cr 
\gamma_{N1} & \gamma_{N2} & \cdots & \cdots & \gamma_{NN} }
$$
We identify 
$$
\gamma_{Nj} = \lambda_j, \quad j = 1 \cdots N
$$
Then
\begin{eqnarray}
w(\lambda,q) &=& \int \prod_{i,j=1}^{N-1}d \gamma_{ij}\; e^{\left( -{1\over i\hbar} \sum_{n,k =1}^N 
q_n (\gamma_{nk}-\gamma_{n-1,k} ) \right)}\times  \label{MellinBarnes} \\
&&\hskip 1cm \prod_{n=1}^{N-1}
\left\{ {\prod_{j=1}^n \prod_{k=1}^{n+1} \hbar^{\gamma_{nj}-\gamma_{n+1,k} \over 
i\hbar} \Gamma({\gamma_{nj}-\gamma_{n+1,k} \over 
i\hbar})  \over \prod_{j<k =1}^n \vert \Gamma({\gamma_{nj}-\gamma_{n,k} \over 
i\hbar})\vert^2 }\right\}\nonumber
\end{eqnarray}
The integration contour is as follows.
\begin{eqnarray*}
{\rm Im}\; \gamma_{11} &>& {\rm max} ({\rm Im}\; \gamma_{12}, {\rm Im}\; \gamma_{22})\\
{\rm Im}\; \gamma_{21}, {\rm Im}\; \gamma_{22} &>& 
 {\rm max} ({\rm Im}\; \gamma_{31}, {\rm Im}\; \gamma_{32},  {\rm Im}\; \gamma_{33})\\
 &\vdots & \\
 {\rm Im}\; \gamma_{N-1,1},\cdots, {\rm Im}\; \gamma_{N-1,N-1} &>&
 {\rm max} ({\rm Im}\; \gamma_{N1}, \cdots,  {\rm Im}\; \gamma_{NN})
\end{eqnarray*}
Note that the poles of the integrand are located at
$$
\gamma_{nj} = \gamma_{n+1,k} - i\hbar s, \quad s \in N  
$$ 
so we can move the contour $\gamma_{nj}$ upward safely.
\end{itemize}

\section{Diagonalization of $X^{(k)}$.}

Since the operators $X^{(k)}$ are all commuting, we can diagonalize 
them simultaneously. The $X^{(k)}$ are self-adjoint with respect 
to the measure $m(\lambda)$ eq.(\ref{measure}), their eigenfunctions satisfy
the same completeness and orthogonality conditions as 
 the Whittaker functions diagonalizing the Hamiltonians of the
open Toda chain eqs.(\ref{completness},\ref{orthogonality}). 

The Whittaker functions are the kernel of the Fourier transform
going from the variables $q_i$ to the variables $\lambda_i$.
Similarly, the eigenfunctions of the opertors $X^{(k)}$ are the 
kernel of the Fourier transform going from the variables $\lambda_i$ to the variables $q_i$. So, it is natural to expect
\begin{equation}
X^{(k)} (\lambda,\mu )\; {w}^*(\lambda, q) = (-1)^{k(N-1)} e^{\sum_{i=1}^k  q_{N+1-i} } \; {w}^*(\lambda, q)
\label{xkeq2}
\end{equation}
Taking the complex conjugate of this equation, we get
\begin{equation}
X^{(k)*} (\lambda,\mu )\; {w}(\lambda, q) = (-1)^{k(N-1)} e^{\sum_{i=1}^k  q_{N+1-i} } \; {w}(\lambda, q)
\label{xkeq3}
\end{equation}
where $X^{(k)*}$ is the complex conjugate of $X^{(k)}$, i.e. 
it is given by eq.(\ref{defXr}) with $\mu_i$ replaced by $\mu_i^*$
 the shift operator $\lambda_i \to \lambda_i-i\hbar$. 
Note that eqs.(\ref{kostant}) are differential 
equations in $q_i$, while eqs.(\ref{xkeq3}) are difference equations 
in $\lambda_i$.

In the remaining of this section, we prove that the function
$w(\lambda,q)$ defined by the Mellin-Barnes integral representation
 eq.(\ref{MellinBarnes}),
which is known to satisfy eqs.(\ref{kostant}), also satisfy eqs.(\ref{xkeq3}).

Before treating the general case, it is instructive to do the calculation for
$N=2$ and $N=3$ first.

\subsection{The case $N=2$.}

For $N=2$, the function $w(\lambda, q)$ reads:
\begin{eqnarray*}
w(\lambda,q) &=& \int d \gamma_{11}\; e^{ -{1\over i\hbar} [ 
q_1 \gamma_{11} + q_2(\lambda_{1} -\gamma_{11} + \lambda_{2} )]}
  \prod_{k=1}^{2}(i \hbar)^{\gamma_{11}-\lambda_{k} \over 
i\hbar} \Gamma\left({\gamma_{11}-\lambda_{k} \over 
i\hbar}\right) 
\end{eqnarray*}
The operators $X^{(k)*}$ are
$$
 X^{(1)^*}= {1\over \lambda_1 - \lambda_2} (\mu_1^* - \mu_2^*) ,\quad
 X^{(2)^*} = \mu_1^* \mu_2^*  
$$
We have
\begin{eqnarray*}
X^{(1)^*} w(\lambda, q) &=& \\
&& \hskip -2.5cm
{i\hbar e^{q_2}\over \lambda_{12}} \int d \gamma_{11}\; e^{ -{1\over i\hbar} [ 
q_1 \gamma_{11} + q_2(\lambda_{1} -\gamma_{11} + \lambda_{2} ) ]}
  \prod_{k=1}^{2} (i\hbar)^{\gamma_{11}-\lambda_{k} \over 
i\hbar} \Gamma\left({\gamma_{11}-\lambda_{k} \over 
i\hbar}\right) \left[{(\gamma_{11} - \lambda_1)\over i\hbar} -{ (\gamma_{11} - \lambda_2)\over i\hbar}\right] = \\
&& 
- e^{q_2}\int d \gamma_{11}\; e^{ -{1\over i\hbar} [
q_1 \gamma_{11} + q_2(\lambda_{1} -\gamma_{11} + \lambda_{2} ) ]}
  \prod_{k=1}^{2} (i\hbar)^{\gamma_{11}-\lambda_{k} \over 
i\hbar} \Gamma\left({\gamma_{11}-\lambda_{k} \over 
i\hbar}\right)
\end{eqnarray*}
Hence, we have proved 
$$
X^{(1)*} w(\lambda,q)= -e^{q_2} w(\lambda,q)
$$
Similarly
\begin{eqnarray*}
X^{(2)*} w(\lambda,q) 
&=& \int d \gamma_{11}\; e^{ -{1\over i\hbar} [ 
q_1 \gamma_{11} + q_2(\lambda_{1} -\gamma_{11} + \lambda_{2}-2i\hbar ) ]}
  \prod_{k=1}^{2} (i\hbar)^{\gamma_{11}-\lambda_{k} +i\hbar\over 
i\hbar} \Gamma\left({\gamma_{11}-\lambda_{k} +i\hbar \over 
i\hbar}\right) = \\
&=& 
 e^{q_2 + q_1}\int d \gamma_{11}'\; e^{ -{1\over i\hbar} [
q_1 \gamma_{11}' + q_2(\lambda_{1} -\gamma_{11}' + \lambda_{2} ) ]}
  \prod_{k=1}^{2} (i\hbar)^{\gamma_{11}'-\lambda_{k} \over 
i\hbar} \Gamma\left({\gamma_{11}'-\lambda_{k} \over 
i\hbar}\right)
\end{eqnarray*}
where $\gamma_{11}' = \gamma_{11} +i\hbar$. So, the integral on the right hand side is the
same as the initial one but on a contour shifted upward by $+i\hbar$. But as we already noticed, this does not change
the value of the integral. Hence, we have proved
$$
X^{(2)*} w(\lambda,q)= e^{q_1+q_2} w(\lambda,q)
$$

\subsection{The case $N=3$.}

For $N=3$, we have:
\begin{eqnarray}
w(\lambda,q) &=& \int d \gamma\; e^{ -{1\over i\hbar} [ 
q_1 \gamma_{11} + q_2(\gamma_{21} -\gamma_{11} + \gamma_{22}) 
+ q_3(\lambda_1 +\lambda_2 +\lambda_3 - \gamma_{21}-\gamma_{22} ) ] }\times  \label{MB3}\\
&& \hskip 1cm \times {\prod_{j,k} (i\hbar)^{\gamma_{1j}-\gamma_{2k}\over i\hbar}    
\Gamma\left({\gamma_{1j}-\gamma_{2k} \over i\hbar} \right)
 \over 
\prod_{j<k}
\vert \Gamma\left( {\gamma_{1j}-\gamma_{1k}\over i\hbar}\right)\vert^2
}
\times
{
(i\hbar)^{\gamma_{2j}-\lambda_{k}\over i\hbar}    
\Gamma\left({\gamma_{2j}-\lambda_{k}\over  i\hbar}\right) \over 
\prod_{j<k}
\vert \Gamma\left( {\gamma_{2j}-\gamma_{2k}\over i\hbar}\right)\vert^2
  }\nonumber
\end{eqnarray}
The operators $X^{(k)*}$ read 
\begin{eqnarray*}
 X^{(1)*}&=& {1\over \lambda_{12}\lambda_{13}} \mu_1^*
+{1\over \lambda_{21}\lambda_{23}} \mu_2^*
+{1\over \lambda_{31}\lambda_{32}} \mu_3^* \\
 X^{(2)*} &=&  
{1\over \lambda_{13} \lambda_{23}} \mu_1^*\mu_2^*
+{1\over \lambda_{12}\lambda_{32}} \mu_1^*\mu_3^*
+{1\over \lambda_{21}\lambda_{31}} \mu_2^*\mu_3^* \\
X^{(3)*} &=&  \mu_1^*\mu_2^*\mu_3^*
\end{eqnarray*}

Consider first  $X^{(1)*} w(\lambda,q)$. The exponential factor in the first line of eq.(\ref{MB3}) produces 
the factor $e^{q_3}$. In the integrand, using $\Gamma(x+1)=x\Gamma(x)$, we get the factor
$$
 {(i\hbar)^2\over \lambda_{12}\lambda_{13}} {(\gamma_{21}-\lambda_1)\over i\hbar}{(\gamma_{22}-\lambda_1)\over i\hbar}
+{(i\hbar)^2\over \lambda_{21}\lambda_{23}}{ (\gamma_{21}-\lambda_2)\over i\hbar}{(\gamma_{22}-\lambda_2)\over i\hbar}
+{(i\hbar)^2\over \lambda_{31}\lambda_{32}} {(\gamma_{21}-\lambda_3)\over i\hbar}{(\gamma_{22}-\lambda_3)\over i\hbar} 
$$
which is equal to $1$.  Hence, we have proved
$$
X^{(1)*} w(\lambda,q) =  e^{q_3} w(\lambda,q)
$$

Next, we consider $X^{(2)*}w(\lambda,q)$. The exponential factor produces the factor $e^{2q_3}$. Using the 
$\Gamma$ function relation, we get in the integrand
\begin{eqnarray*}
{1\over \lambda_{13} \lambda_{23}} (\gamma_{21}-\lambda_1)(\gamma_{22}-\lambda_1)
 (\gamma_{21}-\lambda_2)(\gamma_{22}-\lambda_2)  &+& \\
{1\over \lambda_{12}\lambda_{32}} (\gamma_{21}-\lambda_1)(\gamma_{22}-\lambda_1)
 (\gamma_{21}-\lambda_3)(\gamma_{22}-\lambda_3)  &+& \\
{1\over \lambda_{21}\lambda_{31}}  (\gamma_{21}-\lambda_2)(\gamma_{22}-\lambda_2)
 (\gamma_{21}-\lambda_3)(\gamma_{22}-\lambda_3) &&
\end{eqnarray*}
which is equal to
$$
{(\gamma_{21}-\lambda_1)(\gamma_{21}-\lambda_2)(\gamma_{21}-\lambda_3)-
(\gamma_{22}-\lambda_1)(\gamma_{22}-\lambda_2)(\gamma_{22}-\lambda_3)
\over (\gamma_{21} -\gamma_{22})}
$$
So, we get a sum of two identical  integrals but with $\gamma_{21}$ and $\gamma_{22}$ interchanged.
Let us treat the first one. Using the $\Gamma$ functions, we reconstruct
\begin{eqnarray*}   
\Gamma\left({\gamma_{11}-\gamma_{2k} \over i\hbar} \right)
 \times
{\prod_{k}
\Gamma\left({\gamma_{21}-\lambda_{k}\over  i\hbar}+1\right)\Gamma\left({\gamma_{22}-\lambda_{k}\over  i\hbar}\right) \over 
 \Gamma\left( {\gamma_{21}-\gamma_{22}\over i\hbar}+1\right) \Gamma\left( {\gamma_{22}-\gamma_{21}\over i\hbar}\right)
  }
\end{eqnarray*}
We now change variables $\gamma_{21} = \gamma_{21}' -i\hbar$. The exponential produces the factor
$e^{q_2-q_3}$. The integrand becomes
\begin{eqnarray*}    
\Gamma\left({\gamma_{11}-\gamma_{21} \over i\hbar} \right)\Gamma\left({\gamma_{11}-\gamma_{22} \over i\hbar} \right)
 \times
{\prod_{k}    
\Gamma\left({\gamma_{21}-\lambda_{k}\over  i\hbar}\right)\Gamma\left({\gamma_{22}-\lambda_{k}\over  i\hbar}\right) \over 
 \Gamma\left( {\gamma_{21}-\gamma_{22}\over i\hbar}\right) \Gamma\left( {\gamma_{22}-\gamma_{21}\over i\hbar}\right)
  }\times    {\gamma_{11} - \gamma_{21}\over \gamma_{22}-\gamma_{21}} 
\end{eqnarray*}
Note that there is no pole at $\gamma_{21}=\gamma_{11}$ nor at
$\gamma_{21}=\gamma_{22}$, so that we can move back the $\gamma_{12}$ contour to its original position.
To this, we have to add the same expression with $\gamma_{21}$ and $\gamma_{22}$ interchanged. We
reproduce $w(\lambda,q)$ because
$$
{\gamma_{11} - \gamma_{21}\over \gamma_{22}-\gamma_{21}} + {\gamma_{11} - \gamma_{22}\over \gamma_{21}-\gamma_{22}} =1
$$

Thus we have proved
$$
X^{(2)*}w(\lambda,q) = e^{q_2+q_3} w(\lambda,q)
$$

Finally, we consider $X^{(3)*}w(\lambda,q)$. The exponential factor produces the factor $e^{3q_3}$. Then we change
variables $\gamma_{2j} +i\hbar = \gamma_{2j}'$, $\gamma_{11} +i\hbar = \gamma_{11}'$, which amounts
to shifting the contours by $i\hbar$ upward. This produces a factor $e^{q_1+q_2-2q_3}$.  So we have 
proved that
$$
X^{(3)*} w(\lambda,q) = e^{q_1 + q_2 + q_3} w(\lambda,q)
$$

\subsection{General case.}
We have
$$
X^{(k)*} w(\lambda,q) = \sum_{[k]} S_{[k]} \mu_{[k]}^* w(\lambda,q)
$$
The exponential in the first line of eq.(\ref{MellinBarnes}) produces a factor $e^{kq_N}$. In the integrand
we get the factor
$$
\sum_{[k]}S_{[k]}(\gamma_N)\prod_{j\in [k]}\prod_{i=1}^{N-1} (\gamma_{N-1,i}-\gamma_{Nj})
$$
By eq.(\ref{NtoN-1}) in the Appendix, this is equal to
$$
(-1)^{(N-k)k}
\sum_{[k-1]}S_{[k-1]}(\gamma_{N-1}) \prod_{i\in [k-1]}\prod_{j=1}^{N} (\gamma_{N-1,i}-\gamma_{Nj})
$$
Let us keep track of 
 $\gamma_{N-1,i}, i \in [k-1]$. The factors $\prod_{i\in [k-1]} \prod_{j=1}^N 
 (\gamma_{N-1,i} - \gamma_{N,j})$ are absorbed into the $\Gamma$-functions
 in the numerator of the integrand, while the factor $S_{[k-1]}(\gamma_{N-1})$ is absorbed in the 
 $\Gamma$-functions in the denominator of the integrand,  to produce
 $$
 (-1)^{(N-k)k}
 \prod_{i\in [k-1]} 
 {\prod_{j=1}^N  \Gamma\left({\gamma_{N-1,i}-\gamma_{Nj} \over i\hbar} +1 \right)\over
 \prod_{ j\not\in [k-1]}  \Gamma\left({\gamma_{N-1,i}-\gamma_{N-1,j} \over i\hbar} +1 \right)\Gamma\left({\gamma_{N-1,j}-\gamma_{N-1,i} \over i\hbar} \right)}
 $$
Next, we change variables $\gamma_{N-1,i}+i\hbar = \gamma_{N-1,i}'$,
$i\in [k-1]$. The exponential
factor in the first line of eq.(\ref{MellinBarnes}) is symmetrical in all the $\gamma_{N-1,i}$ and yields  
$$
e^{(k-1)(q_{N-1} -q_{N})}
$$ 
Then, the $\Gamma$-functions with shifted arguments are
 $$
 (-1)^{k(N-k)}
 \prod_{i\in [k-1]} 
 {\prod_{j=1}^{N-2}  \Gamma\left({\gamma_{N-2,j}-\gamma_{N-1,i} \over i\hbar} +1 \right)\over
 \prod_{ j\not\in [k-1]} \Gamma\left({\gamma_{N-1,j}-\gamma_{N-1,i} \over i\hbar}+1 \right)}
 $$
 which produce a factor
 $$
 (-1)^{(N-k)}S_{[k-1]}(\gamma_{N-1}) \prod_{i\in [k-1]}\prod_{j=1}^{N-2}
  (\gamma_{N-2,j}-\gamma_{N-1,i} )
 $$
 in the integrand.
 So, we are back to the same problem, but at level $k-1, N-1$. We eventually reach the
 level $1, N-k+1$, where we use the identity
 $$
 \sum_i \prod_{j\neq i} {1\over \gamma_{N-k+1,i}-\gamma_{N-k+1,j}} 
 \prod_{j=1}^{N-k} (\gamma_{N-k,j} - \gamma_{N-k+1,i}) = (-1)^{(N-k)}
 $$ 
 Putting everything together, we arrive at
 $$
 X^{(k)*} w(\lambda,q) = (-1)^{k(N-1)}
 e^{ q_N + q_{N-1}+\cdots + q_{N-k+1}} w(\lambda,q)
 $$
which is exactly eq.(\ref{xkeq3}).

\section{Conclusion.}

To conclude, we would like to mention that
the operators $X^{(k)}$ are limiting cases of the Ruijsenaars-Macdonald 
operators:
$$
M^{(k)} =\sum_{[k]} \prod_{i\in [k] \atop j \not \in [k]}
{t q^{\lambda_i} - t^{-1} q^{\lambda_j} \over 
 q^{\lambda_i} -  q^{\lambda_j} } \mu_{[k]}
$$
So, our result is probably a limiting case of the results obtained
in \cite{ESV02}.

{\bf Acknowledgements:} I thank F. Smirnov for discussions.
This work was partially supported by the European Commission 
TMR program HPRN-CT-2002-00325 (EUCLID).

\section{Appendix.}
\begin{proposition}
One has the identity
\begin{equation}
\sum_{[k]}S_{[k]}(\gamma_N)\prod_{j\in [k]}\prod_{i=1}^{N-1} (\gamma_{N-1,i}-\gamma_{Nj}) = (-1)^{(N-k)k}
\sum_{[k-1]}S_{[k-1]}(\gamma_{N-1}) \prod_{i\in [k-1]}\prod_{j=1}^{N} (\gamma_{N-1,i}-\gamma_{Nj})
\label{NtoN-1}
\end{equation}
\end{proposition}
\proof
First of all, the left-hand side is a polynomial in $\gamma_N$. It has potential poles at $\gamma_{Ni}=\gamma_{Nj}$.
Let us suppose $i=1$, $j=2$. One has to assume that $1 \in [k], 2 \not\in [k]$ or vice versa, otherwise there
is no pole. In the above sum, we consider the two terms 
$$
[k] = 1 + [k'],\quad [k] = 2 + [k'],\quad 1,2 \not\in [k']
$$
Denote by $[l']$ the complementary subset of $[k']$ in $3,4,\cdots, N$. The two terms are, respectively
$$
{1\over \gamma_{N1}-\gamma_{N2}}\prod_{j\in [l']}{1\over \gamma_{N1}-\gamma_{Nj}}
\prod_{i\in [k']}{1\over \gamma_{Ni}-\gamma_{N2}}\prod_{i\in [k']\atop j \in [l']}
{1\over \gamma_{Ni}-\gamma_{Nj}} \prod_{j=1}^{N-1}(\gamma_{N-1,j} - \gamma_{N1})
 \prod_{i\in [k']}(\gamma_{N-1,j} - \gamma_{Ni})
 $$
$$
{1\over \gamma_{N2}-\gamma_{N1}}\prod_{j\in [l']}{1\over \gamma_{N2}-\gamma_{Nj}}
\prod_{i\in [k']}{1\over \gamma_{Ni}-\gamma_{N1}}\prod_{i\in [k']\atop j \in [l']}
{1\over \gamma_{Ni}-\gamma_{Nj}} \prod_{j=1}^{N-1}(\gamma_{N-1,j} - \gamma_{N2})
 \prod_{i\in [k']}(\gamma_{N-1,j} - \gamma_{Ni})
 $$
It follows that the sum of the two residues cancel. From the behaviour at $\infty$, we see that
both sides are polynomials of degree $k-1$. To show that they are identical, we compare the values
at the $N-1$ points $\gamma_{N1} = \gamma_{N-1,i}$. It is enough to consider 
$$
\gamma_{N1} = \gamma_{N-1,1}
$$
In the left hand side, only the sets $[k]$ such that $1\not\in [k]$ contribute. Hence we get
$$
\sum_{1\not\in [k]}\prod_{i\in [k]} {1\over \gamma_{Ni}-\gamma_{N1}} \prod_{i\in [k] \atop
j\not\in [k], j\neq 1}{1\over \gamma_{Ni}-\gamma_{Nj}} \prod_{j\in [k]} (\gamma_{N-1,1}-\gamma_{Nj})
\prod_{i=2}^{N-1} (\gamma_{N-1,i}-\gamma_{Nj})
$$
When evaluated at $\gamma_{N1} = \gamma_{N-1,1}$, the first and third terms cancel and we are left with
$$
(-1)^k \sum_{1\not\in [k]} \prod_{i\in [k] \atop
j\not\in [k], j\neq 1}{1\over \gamma_{Ni}-\gamma_{Nj}} 
\prod_{i=2}^{N-1}\prod_{j\in [k]} (\gamma_{N-1,i}-\gamma_{Nj})
$$
In the right hand side, only the sets $[k-1]$ such that $1\not\in [k-1]$ contribute. Hence, we get
$$
(-1)^{(N-k)k}\sum_{[k-1]}\prod_{i\in [k-1] }{1\over \gamma_{N-1,i}-\gamma_{N-1,1}}
\prod_{i\in [k-1] \atop j\neq 1}{1\over \gamma_{N-1,i}-\gamma_{N-1,j}}
 \prod_{i\in [k-1]}(\gamma_{N-1,i}-\gamma_{N1})\prod_{j=2}^{N} (\gamma_{N-1,i}-\gamma_{Nj})
$$
When evaluated at $\gamma_{N1} = \gamma_{N-1,1}$, the first and third terms cancel and we are left with
$$
(-1)^{(N-k)k}\sum_{1\not\in [k-1]}
\prod_{i\in [k-1] \atop j\neq 1}{1\over \gamma_{N-1,i}-\gamma_{N-1,j}}
 \prod_{i\in [k-1]}\prod_{j=2}^{N} (\gamma_{N-1,i}-\gamma_{Nj})
$$
The two things are identical if our identity holds at level $N-1$. The lowest level is $N=k$. There, we have
$$
S_{[N]}(\gamma_N) = 1,\quad S_{[N-1]}(\gamma_{N-1}) = 1
$$
and the identity reduces to
$$
\prod_{j=1}^N \prod_{i=1}^{N-1}(\gamma_{N-1,i}-\gamma_{Nj}) =
 \prod_{i=1}^{N-1}\prod_{j=1}^N(\gamma_{N-1,i}-\gamma_{Nj})
 $$
 which is obviously true. \square

 \end{document}